\documentclass[12pt]{article}
\pdfoutput=1
\usepackage{latexsym}
\usepackage{cite,epsfig,amssymb,euscript,slashed}
\usepackage[normalem]{ulem} %defines \sout{word} which strikes out {word}
\usepackage{amsmath}
\usepackage{float}
\usepackage{bbm}
\usepackage{braket}
\usepackage{xcolor}
\usepackage{fancybox}
\usepackage{geometry}
\if{}
\fi

\def\be{\begin{equation}}
\def\ee{\end{equation}}
\def\bseq{\begin{subequations}}
\def\eseq{\end{subequations}}

\def\bea{\begin{eqnarray}}
\def\eea{\end{eqnarray}}

\def\bseq{\begin{subequations}}
\def\eseq{\end{subequations}}

\numberwithin{equation}{section} %%
\usepackage[]{graphicx}

\usepackage[linktocpage]{hyperref}

\def\ii           {{\rm i}}

\def\sqr#1#2{{\vcenter{\vbox{\hrule height.#2pt
 \hbox{\vrule width.#2pt height#1pt \kern#1pt \vrule width.#2pt}\hrule
 height.#2pt}}}}

\def\m{\mu}
\def\n{\nu}

\def\m{\mu}

\def\n{\nu}

\def\slashchar#1{\setbox0=\hbox{$#1$}           % set a box for #1
\dimen0=\wd0                                 % and get its size
\setbox1=\hbox{/} \dimen1=\wd1               % get siste of /
\ifdim\dimen0>\dimen1                        % #1 is bigger
\rlap{\hbox to \dimen0{\hfil/\hfil}}      % so center / in box
#1                                        % and print #1
\else                                        % / is bigger
\rlap{\hbox to \dimen1{\hfil$#1$\hfil}}   % so center #1
/                                         % and print /
\fi}

\begin{document}
\font\cmss=cmss10 \font\cmsss=cmss10 at 7pt

%\title{ModMax Electrodynamics and Charged Particles}

\title{Maximally symmetric nonlinear extension of electrodynamics and charged particles}

\author{ Kurt Lechner$^{a,b}$, Pieralberto Marchetti$^{a,b}$, Andrea Sainaghi$^a$ and Dmitri P. Sorokin$^{b,a}$}

\date{}

\maketitle

\vspace{-1.5cm}

\begin{center}

\vspace{0.5cm}
\textit{$^a$Dipartimento di Fisica e Astronomia ``Galileo Galilei",  Universit\`a degli Studi di Padova\\
and
\\
$^b$I.N.F.N. Sezione di Padova, Via F. Marzolo 8, 35131 Padova, Italy}
\end{center}

\vspace{5pt}

\abstract{We consider couplings of electrically and magnetically charged sources to the maximally symmetric non-linear extension of Maxwell's theory called ModMax. The aim is to reveal physical effects which distinguish ModMax from Maxwell's electrodynamics. We find that, in contrast to generic models of non-linear electrodynamics, Lienard-Wiechert fields induced by a moving electric or magnetic particle, or a dyon are exact solutions of the ModMax equations of motion.  We then study whether and how ModMax non-linearity affects properties of electromagnetic interactions of charged objects, in particular the Lorentz force, the Coulomb law, the Lienard-Wiechert fields, Dirac's and Schwinger's quantization of electric and magnetic charges, and the Compton Effect. In passing we also present an alternative form of the ModMax Lagrangian in terms of the coupling of Maxwell's theory to axion-dilaton-like auxiliary scalar fields which may be relevant for revealing the effective field theory origin of ModMax.
}
\noindent

%\noindent {\em Possible comment ............  }

\thispagestyle{empty}

%\vfill
%\vskip 5.mm
%\hrule width 5.cm
%\vskip 2.mm
%{\scriptsize
%\noindent e-mails: {\tt sorokin@pd.infn.it
%}}

\newpage

\setcounter{footnote}{0}

\tableofcontents

\newpage

\section{Introduction}
In \cite{Bandos:2020jsw} there was discovered a non-linear extension of free $D=4$ Maxwell's electrodynamics, called ModMax, which preserves all the symmetries of the latter, namely the four-dimensional conformal symmetry and electric-magnetic duality. This is the unique specimen with such properties among the variety of models of non-linear electrodynamics (see \cite{Sorokin:2021tge} for a review). This theory exhibits interesting features, such as a peculiar form of birefringence \cite{Bandos:2020jsw,Flores-Alfonso:2020euz} and, in spite of its intrinsic non-analyticity, has plane waves \cite{Bandos:2020jsw} and topologically non-trivial knotted null electromagnetic field configurations \cite{Dassy:2021ulu} as exact solutions in its Hamiltonian formulation. Properties of the ModMax Lagrangian formulation were studied in more detail in \cite{Kosyakov:2020wxv,Kosyakov:2022klk} and its Hamiltonian formulation in \cite{Escobar:2021mpx}.

ModMax arises as a weak field limit of a generalized two-parameter Born-Infeld theory \cite{Bandos:2020jsw,Bandos:2020hgy}. Further generalizations of these theories were discussed in \cite{Sokolov:2021qfs,Kruglov:2021bhs}. Quite remarkably, as was shown recently \cite{Babaei-Aghbolagh:2022uij,Ferko:2022iru,Conti:2022egv}, both, ModMax and the generalized Born-Infeld theory arise as different $T\bar T$-like deformations of Maxwell's and the Born-Infeld theory, and the generalized Born-Infeld theory is a $T\bar T$ deformation of ModMax. Their supersymmetric extensions were constructed in \cite{Kuzenko:2021cvx,Bandos:2021rqy}, and in \cite{Kuzenko:2021qcx} (super)conformal higher-spin generalizations of ModMax were derived. A possibility of linking the generalized Born-Infeld theory to String Theory by uncovering a string-like nature of the former was discussed in \cite{Nastase:2021uvc}.

Effects of ModMax and its generalizations on properties and thermodynamics of charged black holes (e.g. Taub-NUT, Reissner-Nordstr\"om ones and others) have been studied in a number of papers \cite{Flores-Alfonso:2020euz,Bordo:2020aoz,Flores-Alfonso:2020nnd,Amirabi:2020mzv,Bokulic:2021dtz,Zhang:2021qga,Bokulic:2021xom,Ali:2022yys,Kruglov:2022qag,Ortaggio:2022ixh,Barrientos:2022bzm}.
 The aim of this article is to study how the ModMax non-linearity affects properties of interactions of electrically and magnetically charged point particles, in particular the Lorentz force, the Coulomb law, the Lienard-Wiechert fields, Dirac's and Schwinger's quantization of electric and magnetic charges, and the Compton effect. We will show that the Lienard-Wiechert fields created by a moving electric or magnetic particle, or a dyon are exact solutions of the ModMax equations of motion, while this is not the case for most of non-linear electrodynamics models. In passing we will also present an alternative form of the ModMax Lagrangian in terms of the coupling of Maxwell's theory to axion-dilaton-like auxiliary scalar fields which may be relevant for revealing the effective field theory origin of ModMax.

 We will show that for a certain choice of the definition of physical electric and magnetic charges, which is associated with an appropriate rescaling of the source-free ModMax Lagrangian, and the standard minimal electromagnetic coupling of the charges, there is no difference in the Coulomb law and Lorentz forces describing interactions of two electric particles (one of which is a test particle) in ModMax and Maxwell's theory. The difference appears if magnetic monopoles or dyons are present. On the other hand, as is the case of vacuum birefringence of ModMax \cite{Bandos:2020jsw,Flores-Alfonso:2020euz}, the Compton scattering differs from that in Maxwell's theory for any scaling of the ModMax Lagrangian.

 {\bf Notation and conventions.} We use the almost minus signature of the Lorentz metric $(+,-,-,-)$ and natural units in which the speed of light $c$ and the Planck constant $\hbar$ are set to one.

 \section{ModMax electrodynamics}\label{MME}
The Lagrangian density of ModMax has the following form \cite{Bandos:2020jsw}
\begin{eqnarray}\label{MML}
{\cal L}& = & {\cosh\gamma} \,S + {\sinh\gamma}\, \sqrt{S^2+P^2}\, \\
%&= &-\frac {\cosh\gamma} 4\, F_{\mu\nu}F^{\mu\nu}+ \frac{\sinh\gamma}4\, \sqrt{(F_{\mu\nu}F^{\mu\nu})^2+(F_{\mu\nu}\tilde F^{\mu\nu})^2}\, \\
&&\nonumber\\
&= &\frac {\cosh\gamma} 2\, (\mathbf E^2-\mathbf B^2)+ \frac{\sinh\gamma}2\, \sqrt{(\mathbf E^2-\mathbf B^2)^2+4(\mathbf E\cdot \mathbf B)^2}\,,\nonumber
\end{eqnarray}
where
\be\label{SP}
S=-\frac 14 F_{\mu\nu}F^{\mu\nu}=\frac 12 \,({\mathbf E}^2-{\mathbf B}^2)\,,\qquad P=-\frac 14 F_{\mu\nu}\tilde F^{\mu\nu}={\mathbf E}\cdot{\mathbf B}
\ee
are the two independent $D=4$ Lorentz invariants constructed from the electromagnetic field strength
$F_{\mu\nu}=\partial_\mu A_\nu-\partial_\nu A_\mu$ and its Hodge dual $\tilde F^{\mu\nu}=\frac 12\varepsilon^{\mu\nu\rho\lambda}F_{\rho\lambda}$. $E_i=F_{0i}$ is the electric three-vector field ($i=1,2,3$), $B^i=\tilde F^{0i}$ is the magnetic vector field and
 $\gamma$ is a dimensionless coupling constant. The conditions of causality and unitarity require this constant to be non-negative $\gamma \geq 0$ \cite{Bandos:2020jsw}. These values of $\gamma$ also ensure that the Lagrangian density is a convex function of the electric field $E_i$ \cite{Bandos:2020jsw} and that its energy-momentum tensor satisfies the weak, strong and dominant energy conditions \cite{Pergola:2021}. Note that Maxwell's electrodynamics is not a weak field limit of ModMax because of conformal invariance, but is recovered when the ModMax coupling constant tends to zero $\gamma\to 0$.

The Lagrangian field equations of the theory, accompanied by the Bianchi identities, are
\begin{equation}\label{MMLeom}
\partial_\mu G^{\mu\nu} =\cosh\gamma\,\partial_\mu F^{\mu\nu} +\sinh\gamma\,\partial_\mu\,\Bigl(\frac{SF^{\mu\nu} + P\Tilde{F}^{\mu\nu}}{\sqrt{S^{2} + P^{2}}}\Bigr)=0,\qquad \partial_\mu \tilde F^{\mu\nu}=0,
\end{equation}
where
\be\label{G}
G^{\mu\nu}:=-2\frac{\partial \mathcal L}{\partial F_{\mu\nu}}\,=\cosh\gamma\, F^{\mu\nu} +\sinh\gamma\,\Bigl(\frac{SF^{\mu\nu} + P\Tilde{F}^{\mu\nu}}{\sqrt{S^{2} + P^{2}}}\Bigr).
\ee
The equations are non-linear, but they linearize for field configurations for which $P=c\,S$ with $c$ being a constant. So all the solutions of Maxwell's equations with $P=c\,S$ are solutions of ModMax theory.

\if
Since ModMax is conformal, its energy-momentum tensor is traceless and proportional to that of Maxwell's theory
\begin{eqnarray}\label{EMT}
T^{\mu\nu}&=&\left(F^{\mu}{}_{\rho}F^{\nu\rho}-\frac 14 \eta^{\mu\nu} (F_{\rho\lambda}F^{\rho\lambda})\right){\mathcal L_S}\,,\\
\mathcal L_S&=&\frac{\partial\mathcal L}{\partial S}=\left(\cosh\gamma-\sinh\gamma\frac{F_{\sigma\nu}F^{\sigma\nu}}{\sqrt{(FF)^2+(F\tilde F)^2}}\right)\,,\nonumber\\
T^{\mu}{}_{\mu}&=0\,.&\nonumber
\end{eqnarray}
\fi
One can notice that the equations of motion \eqref{MMLeom} are non-analytic and are not well defined when the electromagnetic fields are null, i.e. the fields for which the Lorentz scalar and pseudo-scalar \eqref{SP} are zero
\be\label{null}
S=0,\qquad P=0.
\ee
In the null-field limit the ambiguity of the values of the scalar factors $\frac S{\sqrt{S^{2} + P^{2}}}$ and $\frac P{\sqrt{S^{2} + P^{2}}}$ in \eqref{MMLeom} range from $-1$ to $+1$. This might be an issue, since the class of solutions for which the electromagnetic fields are null, such as the plane waves, are not well defined in the ModMax Lagrangian formulation.\footnote{Note, though, that the vacuum (in which $F_{\mu\nu}=0$) is a well defined solution of the ModMax Lagrangian field equations \eqref{MMLeom} with zero energy-momentum tensor.} However, somewhat surprisingly, the Hamiltonian formulation of ModMax comes to the rescue \cite{Bandos:2020jsw}. In the ModMax Hamiltonian formulation the null electromagnatic fields are well defined. Among these configurations, the plane waves \cite{Bandos:2020jsw} and topologically non-trivial knotted electromagnetic fields \cite{Dassy:2021ulu} (generalizing those of Maxwell's theory \cite{Arrayas:2017sfq}) are exact solutions of the ModMax Hamiltonian equations (see \cite{Bandos:2020jsw,Dassy:2021ulu,Sorokin:2021tge} for more details).

\subsection{Conformal and duality invariance}
The ModMax action $I=\int d^4x \,\mathcal L(S,P)$ is invariant under the $D=4$ conformal transformations \cite{Bandos:2020jsw,Bandos:2020hgy}, which can be easily checked for the rescaling of the coordinates and the fields with a constant \hbox{parameter $b$}
\be\label{dilatation}
x^\mu \to b\, x^\mu, \quad A_\mu \to b^{-1} A_\mu, \quad F_{\mu\nu} \to b^{-2} F_{\mu\nu}.
\ee
The ModMax field equations and the Bianchi identities \eqref{MMLeom} are invariant under electric-magnetic duality $SO(2)$ rotations of $G^{\mu\nu}$ and $\tilde F^{\mu\nu}$ \cite{Bandos:2020jsw,Kosyakov:2020wxv}
\be\label{Duality}
 \begin{pmatrix}G^{\,\mu\nu}(F')\\
 \tilde F'^{\,\mu\nu}\end{pmatrix}
=
\begin{pmatrix} \cos\alpha & \sin\alpha\\
 -\sin\alpha & \cos\alpha \end{pmatrix}\,
\begin{pmatrix}G^{\mu\nu}(F)\\
 \tilde F^{\mu\nu}\end{pmatrix}\,.
\ee
The duality invariance is ensured by the fact that the ModMax Lagrangian density \eqref{MML} satisfies a condition which must hold for any duality-invariant non-linear electrodynamics \cite{BialynickiBirula:1983tx}, namely
\be\label{Dualitycond}
F_{\mu\nu}\tilde F^{\mu\nu}-G_{\mu\nu}\tilde G^{\mu\nu}=0\,\qquad \Rightarrow \qquad {\mathcal L}_S^2 - \frac{2S}{P}{\mathcal L}_S {\cal L}_P - {\mathcal L}_P^2 =1,
\ee
where $\mathcal L_S= \frac{\partial\mathcal L}{\partial S}$ and $\mathcal L_P= \frac{\partial\mathcal L}{\partial P}$.

\section{Alternative forms of the ModMax Lagrangian}
The ModMax Lagrangian density \eqref{MML} contains the square root function of $S$ and $P$, which may hinder the quantization of this theory.
Using three auxiliary scalar fields $\psi_1,\psi_2$ and $\rho$ one can construct a Lagrangian density which is classically equivalent to ModMax and has the following analytic (polynomial) form\footnote{This is somewhat similar to the case of the Born-Infeld theory in which the square-root can be removed by introducing four real auxiliary scalar fields \cite{Rocek:1997hi}.}
\be\label{MM3}
\mathcal L= \cosh \gamma \, S + \sinh \gamma \left(S\psi_1 + P\psi_2\right)-\frac12\,\rho^{2}(\psi_1^2+\psi^2_2-1)\,.
\ee
The equations of motion for the scalar fields are
\be\label{psi12}
\sinh \gamma\, S=\rho^{2}\psi_1\,,\qquad \sinh \gamma \, P=\rho^{2}\psi_2\,
\ee
and
\be\label{lambda}
\rho(\psi_1^2+\psi^2_2-1)=0
\ee
which for $\rho\not =0$ reduces to
\be\label{lambda0}
(\psi_1^2+\psi^2_2-1)=0\,.
\ee
For $\rho=0$, the values of $\psi_1$ and $\psi_2$ are not defined from \eqref{lambda}, while from \eqref{psi12} we get $S=0=P$, i.e. the electromagnetic fields are null. In this case the Lagrangian density \eqref{MM3}
 reduces to
\be\label{BBL}
\mathcal L= (\cosh \gamma \, + \psi_1\sinh \gamma) \,S  +(\sinh \gamma\,\psi_2)\, P\,
\ee
which is nothing but the Lagrangian density of the Bialynicki-Birula theory describing all possible configurations of the null electromagnetic fields (see \cite{BialynickiBirula:1983tx,BialynickiBirula:1992qj, Bandos:2020hgy} for details about this theory).

Let us now proceed with the case of finite $\rho$. Substituting the solution of equations \eqref{psi12} for $\psi_1$ and $\psi_2$ into \eqref{MM3} one gets
\be\label{MMlambda}
\mathcal L= \cosh \gamma \, S + \frac 12\left(\,{\rho}^{-2}\,\sinh^2 \gamma \left(S^2 + P^2\right)+{\rho}^{2}\right).
\ee
Now the equation of motion of $\rho$ gives
\be\label{lambda1}
 \rho^{4}=\sinh^2\gamma \left(S^2 + P^2\right)\,.
\ee
Substituting this back into \eqref{MMlambda} we get the original ModMax Lagrangian density \eqref{MML}.

Note that the choice of the square of the field $\rho$ in the Lagrangian densities \eqref{MM3} and \eqref{MMlambda} ensured that \eqref{lambda1} has the unique solution with the positive sign on its right hand side (provided that $\gamma>0$). If instead of $\rho^{2}$ we chose a generic auxiliary field $\hat\rho$, the corresponding field equation would have two solutions $\hat\rho=\pm \sinh \gamma \sqrt{S^2 + P^2}$, which would correspond to the ModMax Lagrangian densities with the plus and minus sign of $\sinh\gamma$ in \eqref{MML}, respectively.

Alternatively, the equations \eqref{lambda} and \eqref{psi12} can be solved as follows
\be\label{cossin}
\psi_1=\cos\varphi, \qquad \psi_2=\sin\varphi,
\ee
\be\label{>0}
\rho^{2}=\frac{S\,\sinh\gamma}{\cos\varphi}=\frac{P\,\sinh\gamma}{\sin\varphi}>0.
\ee
Substituting the expressions for $\psi_1$ and $\psi_2$ as the functions of the single auxiliary field $\varphi$ into \eqref{MM3}, one gets the Lagrangian density of the following form
\be\label{MMLphi}
{\cal L}= \cosh \gamma \, S + \sinh \gamma \left(S\cos\varphi + P\sin\varphi\right)\; .
\ee
To get back the original ModMax Lagrangian density \eqref{MML} one eliminates the auxiliary field $\varphi$ by solving its equation of motion and substitutes the solution back into eq. \eqref{MMLphi}
\be\label{phieq}
S\,\sin\varphi-P\,\cos\varphi=0 \quad\Rightarrow \quad \tan\varphi=\frac PS\,\quad\Rightarrow \quad \sin\varphi=\pm \frac P{\sqrt{S^2+P^2}}\,,\quad \cos\varphi=\pm \frac S{\sqrt{S^2+P^2}}\,.
\ee
Substituting into \eqref{MMLphi} the solution for $\varphi$ in \eqref{phieq} with the upper sign we recover again the ModMax Lagrangian density \eqref{MML}. In the formulation with the three auxiliary fields, the solution of \eqref{phieq} with the minus sign is excluded by the positive definiteness of the expressions in \eqref{>0}.

If, however, one considers the Lagrangian density \eqref{MMLphi} as an independent starting point without imposing the conditions
\eqref{>0},
then {\it a priori} the second solution in \eqref{phieq} cannot be excluded and, upon substitution into \eqref{MMLphi}, results in the Lagrangian density
\be\label{MML-}
{\cal L} =  {\cosh\gamma} \,S - {\sinh\gamma}\, \sqrt{S^2+P^2}\,.
\ee
For $\gamma>0$ this Lagrangian density has problems with causality and unitarity, but is ``healthy" for $\gamma\leq 0$ and coincides with \eqref{MML} after the replacement $\gamma \to -\gamma$. So, starting with the Lagrangian density \eqref{MMLphi} without requiring \eqref{>0}, upon elimination of the auxiliary field $\varphi$ one gets the ModMax Lagrangian with positive or negative coupling \hbox{constant $\gamma$.}

In spite of the above issue which requires further study, it is interesting to have a closer look at the Lagrangian density \eqref{MMLphi}. Let us note that the coupling between the scalar and electromagnetic fields described by the term
$S\,\cos\varphi+P\,\sin\varphi$ was introduced in \cite{Balakin:2021jby,Balakin:2021arf} as a building block of a model which the authors called non-linear axion-dilaton electrodynamics. Here we would like to discuss a similarity of the Lagrangian density \eqref{MMLphi} to that of Maxwell's theory coupled to an axion $a(x)$ and a dilaton $\phi(x)$
\be\label{Mad}
\mathcal L=-\frac{e^{-\phi}}4 F_{\mu\nu}F^{\mu\nu}+\frac a4 \,F_{\mu\nu}\tilde F^{\mu\nu}\,=e^{-\phi}\,S-a\,P.
\ee
If in \eqref{MMLphi} we formally promote the coupling constant $\gamma$ to a scalar field $\gamma(x)$ we see that the Lagrangians \eqref{MMLphi} and \eqref{Mad} are related to each other by the  field redefinitions
\be\label{ad}
a=-\sinh\gamma\,\sin\varphi\,,\qquad e^{-\phi}=\cosh\gamma+\sinh\gamma\,\cos\varphi\,
\ee
whose inverse are
\be\label{adinverse}
\cosh\gamma=\frac{e^\phi}2(e^{-2\phi}+a^2+1)\,,\qquad \sin\varphi=\mp\frac{2a\,e^{-\phi}}{\sqrt{(e^{-2\phi}+a^2+1)^2-4e^{-2\phi}}}\,.
\ee
As is well known, $a$ and $\phi$ parametrize a hyperbolic space $SL(2,R)/SO(2)$. So eqs. \eqref{ad} and \eqref{adinverse} are just the change of coordinates on $SL(2,R)/SO(2)$ from $(\gamma,\varphi)$ to $(a,\phi)$ and vice versa. More precisely the coordinates $(\gamma,\varphi)$ parametrize two copies of $SL(2,R)/SO(2)$ since the relations \eqref{ad} are invariant under the map $(\gamma,\varphi)\to (-\gamma,\varphi+\pi)$, which is a discrete symmetry of the Lagrangian density \eqref{MMLphi}.

Maxwell's theory coupled to the axion and the dilaton \eqref{Mad} is well known to possess the electric-magnetic duality symmetry which is enhanced from $SO(2)$ to $SL(2, R)$ \cite{Gibbons:1995ap,Gaillard:1997zr}. This symmetry is of course preserved by the field redefinitions \eqref{ad}. If $\gamma$ is constant (as in \eqref{MMLphi}), then $SL(2,R)$ gets broken to $SO(2)$. The infinitesimal $SO(2)$ duality transformation parametrized by $\alpha$ acts on the scalar field $\phi$ as follows
\be\label{so2d}
\delta \varphi=2\alpha\,(\cosh\gamma+\sinh\gamma\,\cos\varphi)=2\alpha\,\mathcal L_S.
\ee
It is obtained from an $SL(2,R)$ transformation of $a$ and $\phi$ which can be found e.g. in \cite{Gibbons:1995ap}. One can notice that $\varphi$ transforms under $SO(2)$ non-linearly and with a constant shift which resembles a Goldstone behaviour of this field.

The axion and the dilaton are dynamical fields whose propagation is described by the $SL(2,R)$ invariant Lagrangian density
\be\label{daL}
\mathcal L_{a,\phi}=\frac 12 \partial_\mu\phi\,\partial^\mu\phi+\frac {e^{2\phi}}2 \partial_\mu\,a\,\partial^\mu\,a\,.
\ee
Substituting the expressions \eqref{ad} into \eqref{daL} one gets the kinetic terms for the fields $\gamma$ and $\varphi$ which have the following form
\be\label{gvL}
L_{\gamma,\varphi}=\frac {
(\cosh 2\gamma+\sinh 2\gamma \cos\varphi)\partial_\mu\gamma\partial^\mu\gamma-2\sinh^2\gamma \sin\varphi\,\partial_\mu\gamma\partial^\mu\varphi+\sinh^2\gamma\,\partial_\mu\varphi\partial^\mu\varphi}{2(\cosh\gamma+\sinh\gamma\,\cos\varphi)^2}\,.
\ee

It would be of interest to figure out whether the Lagrangian density \eqref{MMLphi} with constant $\gamma$ and non-dynamical $\varphi(x)$ can be regarded as a certain conformal effective field theory limit of an axion-dilaton-coupled Maxwell's theory. We hope to address this problem elsewhere.

\section{Coupling of ModMax to electric and magnetic charges}
In this Section we will consider effects of the ModMax non-linearity on the electromagnetic interactions of charged particles.

\subsection{Electrically charged point particles}
Consider first the minimal coupling of ModMax to a point particle of mass $m_e$ carrying an electric charge $e$. The corresponding action has the following form
\begin{equation}
S\left[A,y\right]=\int d^4x\; \mathcal{L}-\int d^4x\;j_e^{\nu}A_{\nu}-m_e\int d\tau \sqrt{\dfrac{dy^{\mu}(\tau)}{d\tau}\dfrac{dy_{\mu}(\tau)}{d\tau}}
\label{MM+e}
\end{equation}
where $\mathcal{L}$ is the ModMax Lagrangian density \eqref{MML}, $y^\mu(\tau)$ is the particle worldline parametrized by the parameter $\tau$ and
\be\label{je}
 j_e^{\mu}=e\int d\tau\; \delta^{(4)}\left(x-y(\tau)\right)\dfrac{dy^{\mu}}{d\tau}\,.
\ee
The electromagnetic field equations \eqref{MMLeom} now acquire a source and take the following form
\be\label{MMLeom+e}
\partial_\mu G^{\mu\nu} =\cosh\gamma\,\partial_\mu F^{\mu\nu} +\sinh\gamma\,\partial_\mu\,\Bigl(\frac{SF^{\mu\nu} + P\Tilde{F}^{\mu\nu}}{\sqrt{S^{2} + P^{2}}}\Bigr)=j_e^\nu,\qquad \partial_\mu \tilde F^{\mu\nu}=0.
\ee
The equations of motion obtained by the variation of \eqref{MM+e} with respect to the particle worldline $y^\mu(\tau)$ describe the conventional Lorentz force acting on the electrically charged particle
\begin{equation}
\dfrac{dp^\mu}{d\tau}=e F^{\mu\nu}(y)\,v_\nu\,,
\label{lorentz force charge}
\end{equation}
where
$$
v^\mu=\frac{d y^\mu(\tau)}{d\tau}, \qquad p^\mu=\frac{m\,v^\mu}{\sqrt{v^\nu v_\nu}}
$$
are respectively the relativistic particle velocity and momentum.

As was mentioned in Section \ref{MME}, the equations \eqref{MMLeom+e} are non-linear in general, however for some classes of fields they reduce to Maxwell's equations (modulo a rescaling). For instance, in the case in which the electric and magnetic fields are orthogonal i.e. satisfy $P=0$ (but $S\neq 0$), the equations become linear
\begin{equation}
    e^{\gamma\frac S{|S|}}\,\partial_{\mu}F^{\mu\nu}=j_e^{\nu}\,,\quad\quad \partial_{\mu}\tilde F^{\mu\nu}=0\,.
    \label{eq modmax with P=0}
\end{equation}
If electric and magnetic fields have the same strength, i.e. $S=0$ (but $P\neq 0$) the equations also linearize
\begin{equation}
    \cosh{\gamma}\,\partial_{\mu}F^{\mu\nu}=j_e^{\nu}\,,\quad\quad \partial_{\mu}\tilde F^{\mu\nu}=0.
\end{equation}
More generically we can study the case in which both $S$ and $P$ are nonzero, but proportional to each other, such that $P=c S$ with $c$ being a constant. Then the equations of motion again reduce to linear ones
\begin{equation}
\left(\cosh{\gamma}+\frac S{|S|}\,\dfrac{\sinh{\gamma}}{\sqrt{1+c^2}}\right)\partial_{\mu}F^{\mu\nu}=j_e^{\nu}\,,\quad\quad \partial_{\mu}\tilde F^{\mu\nu}=0\,.
\end{equation}
The possibility that the equations of motion are linearised for certain configurations of electromagnetic fields is due to the conformal invariance of the theory. This does not happen in the non-conformal non-linear electrodynamics such as the Born-Infeld theory, unless $S$ and $P$ are constant. For latest developments in the construction of solutions of duality-symmetric non-linear electrodynamics models see \cite{Mkrtchyan:2022ulc}.

\subsection{Lienard-Wiechert fields}\label{LWC}
Above we have seen that particular classes of electromagnetic fields satisfy equations of motion which are similar to Maxwell ones, but with different effective coupling constants between the fields and the charges. Therefore, the solutions of Maxwell's equations which describe the fields satisfying $P=c S$ with $c$ being a constant, are also the solutions of ModMax theory. For instance, the solutions of the Maxwell equations
\begin{equation}
    \partial_{\mu}F^{\mu\nu}=j_e^{\nu},\quad\quad \partial_{\mu} \tilde F^{\mu\nu}=0
    \label{LW equations}
\end{equation}
describing the fields generated by a moving electrically charged point particle are called Lienard-Wiechert fields (see e.g. \cite{lechner2018classical}). Their $4$-vector potential is
\begin{equation}
    A_{LW}^{\mu}=\dfrac{e}{4\pi}\dfrac{v^{\mu}}{v^{\nu}l_{\nu}}\bigg\vert_{s=s_0}
    \label{LW potential}
\end{equation}
where $s$ is the particle proper time parameter defined by the relation $ds^2=\eta_{\mu\nu}dy^{\mu}(\tau)dy^{\nu}(\tau)$, $l^{\nu}=x^{\nu}-y^{\nu}(s)$, $v^{\nu}(s)=\dfrac{dy^{\nu}(s)}{ds}$, and $s_0$ is the solution of $l^2=0$ with the condition $x^{0}>y^0(s_0)$. The corresponding Lienard-Wiechert field strength has the following form
\begin{equation}
    F_{LW}^{\mu\nu}=\dfrac{e}{4\pi}\dfrac{1}{(l_{\rho}v^{\rho})^3}\left[l^{\mu}v^{\nu}
    +l^{\mu}l_{\lambda}\left(v^{\lambda}{a}^{\nu}-{a}^{\lambda}v^{\nu}\right)-(\mu\leftrightarrow\nu)\right]\bigg\vert_{s=s_0}
   \equiv  {e} f^{\mu\nu}\,, \qquad \partial_\mu\tilde f^{\mu\nu}=0\,,
    \label{LW fields}
\end{equation}
where ${a}^{\nu}(s)=\dfrac{d^2y^{\nu}(s)}{ds^2}$.

One can see that the Lienard-Wiechert fields satisfy the condition $P=0$, since $f_{\mu\nu}\tilde f^{\mu\nu}=0.$ Note also that
\be\label{ftf}
f^{\mu\nu}f_{\mu\nu}=- \frac{1}{8\pi^2 (v^\mu l_\mu)^4}<0 \qquad \Rightarrow \qquad S>0.
\ee
Therefore, we can easily adopt \eqref{LW fields} to be a solution of the ModMax equations (\ref{eq modmax with P=0}) by making the following rescaling
\begin{equation}\label{LWe}
    F^{\mu\nu}_{MMLW}= e^{-\gamma}\,{e} f^{\mu\nu}\,.
\end{equation}
So a difference of ModMax and Maxwell theory is that the electric charge in the Lienard-Wiechert fields gets effectively rescaled by the ModMax coupling constant. Let us elaborate on this difference in more detail by considering an electric particle at rest. In Maxwell's theory the fields produced by this particle are the Coulomb ones. In the ModMax theory the Coulomb fields have the same form, but with the rescaled electric charge
\begin{equation}
    \vec{E}=e^{-\gamma}\dfrac{e}{4\pi}\dfrac{\vec{r}}{|\vec r|^3},\qquad \vec{B}=\vec{0},
\end{equation}
where $\vec{r}=(x,y,z)$ is the position vector in space with the particle placed at $\vec{r}=0$. Therefore, from \eqref{lorentz force charge} it follows that the Coulomb force acted on a test point particle of charge $q$ at the position $\vec{r}_0$ is given by
\be\label{Coulomb}
\vec{F}=e^{-\gamma}\dfrac{e q}{4\pi}\dfrac{\vec{r_0}}{|\vec r_0|^3},
\ee
so that the permittivity of the vacuum differs from Maxwell's theory by the factor $e^{\gamma}$.

The Coulomb force is tested in experiments  with a very high precision, so for ModMax to be a theory consistent with the experiment for electromagnetic fields in the vacuum its coupling constant $\gamma$ should be very small.\footnote{A bound on the value of $\gamma$ which one extracts from data of the experiment PVLAS \cite{Ejlli:2020yhk} on the vacuum birefringence (discussed in Section \ref{Compton}) is \hbox{$\gamma\leq 3\cdot 10^{-22}$} \cite{Sorokin:2021tge}. }
On the other hand, if one assumes that ModMax may serve as an effective theory for the description of electromagnetic properties of certain materials,
there is {\it a priori} no restrictions on $\gamma$. For instance, in some materials the variation of the permittivity $\epsilon$ and the permeability $\mu$ from their vacuum values can be presumably  modelled by a function of $\gamma$.

In Section \ref{difs} we will moreover
show that the Coulomb law in the ModMax theory can be made exactly the same as in Maxwell's electrodynamcs, i.e. the scale factor $e^{\gamma}$ can be removed from the field equations (\ref{eq modmax with P=0}) by rescaling the source-free ModMax Lagrangian but keeping the minimal coupling to the electric charges intact. This is equivalent to an appropriate rescaling of the electric charge in the minimal coupling term of \eqref{MM+e}. However, $e^\gamma$ factors will re-appear in the magnetic sector of the theory, as we will see.

\subsection{Magnetic monopoles}
Let us now add to the action \eqref{MM+e} terms that describe the coupling of the electromagnetic field to a point particle carrying a magnetic charge $g$ (monopole) and having the mass $m_g$.
To this end we will use the Dirac formulation in which the electromagnetic potential of the monopole is defined everywhere except for a line going from the monopole to infinity (the so-called "Dirac's string") (see e.g. \cite{Blagojevic:1985sh,lechner2018classical} for a review). The presence of the  magnetic current along the particle trajectory $z^\mu(\tau)$
\begin{equation}
     j_g^{\mu}=g\int d\tau\; \delta^{(4)}\left(x-z(\tau)\right)\dfrac{dz^{\mu}}{d\tau}\,
     \label{mag current}
\end{equation}
 modifies the Bianchi identity in \eqref{LW equations} as follows
\begin{equation}
   \partial_{\mu}\tilde F^{\mu\nu}=j_g^{\nu}\,.
   \label{identita bianchi con carica senza riscaling}
\end{equation}
The solution of this equation is
\begin{equation}
    F_{\mu\nu}=\partial_{\mu}A_{\nu}-\partial_{\nu}A_{\mu}-\tilde C_{\mu\nu}\,,
    \label{def of F* for monopoles}
\end{equation}
where $\tilde C_{\mu\nu}$ is the Hodge dual of the antisymmetric tensor distribution
\begin{equation}
    C^{\mu\nu}(x)=-g\iint d\tau d\sigma \left(\dfrac{\partial w^{\mu}}{\partial \tau}\dfrac{\partial w^{\nu}}{\partial \sigma}-\dfrac{\partial w^{\nu}}{\partial \tau}\dfrac{\partial w^{\mu}}{\partial \sigma}\right)\delta^{(4)}\left(x-w(\tau,\sigma)\right)
    \label{stringa dirac1}
\end{equation}
that has its support on the Dirac string worldsheet ${\cal N}$ stemming from the monopole
\begin{equation}
    w^{\mu}(\tau,\sigma)=z^{\mu}(\tau)+u^{\mu}(\tau,\sigma),\qquad u^{\mu}(\tau,0)=0\,.
    \label{def of w string}
\end{equation}
Using Stokes' theorem one can directly check that
\be\label{dCj}
\partial_{\mu}C^{\mu\nu}=j_g^{\nu}\,
\ee
and hence eq. \eqref{identita bianchi con carica senza riscaling} holds.
The action describing the coupling of ModMax to the electric and magnetic charges has the following form
\begin{equation}
I\left[A,y,z,u\right]=\int d^4x\; \mathcal{L}-\int d^4x\;j_e^{\mu}A_{\mu}-m_e\int d\tau \sqrt{\dfrac{dy^{\mu}(\tau)}{d\tau}\dfrac{dy_{\mu}(\tau)}{d\tau}}-m_g\int d\tau \sqrt{\dfrac{dz^{\mu}(\tau)}{d\tau}\dfrac{dz_{\mu}(\tau)}{d\tau}}
    \label{action of mod modmax with monopoles}
\end{equation}
where $\mathcal{L}$ is as in (\ref{MML}) but with $F_{\mu\nu}$ defined in (\ref{def of F* for monopoles}).

The action depends on the electromagnetic vector potential, the trajectories of the charges and the worldsheet of the Dirac string. The latter must however be invisible, i.e. the physical effects should not depend on the string position in space.

We will now show, following \cite{Lechner:1999ga}, that the action \eqref{action of mod modmax with monopoles} changes under a different choice of the string worldsheet by a term proportional to $eg$. The requirement of the independence of the path integral of the quantum theory of the Dirac string  results in Dirac's quantization condition for the electric and magnetic charge. First, let us notice that eq. \eqref{dCj} holds for any string which stems from the monopole. Let us take two strings whose worldsheets ${\cal N}_1$ and ${ \cal N}_2$ described by $C_1^{\mu\nu}$ and $C_2^{\mu\nu}$ do not coincide. Since ${\cal N}_1$ and ${ \cal N}_2$ have the same boundary given by the  worldline of the monopole, the boundary-less worldsheet  ${\cal N}_2- { \cal N}_1$ is the boundary of a three-dimensional manifold ${\cal S}$. Poincar\'e duality \cite{chern2012differentiable} then implies for the associated tensorial $\delta$-functions the relations
\be\label{poinc}
C_2^{\m\n}-C_1^{\m\n}=g\,\partial_\rho D^{\rho\mu\nu}\quad \leftrightarrow  \quad  {\cal N}_2-{\cal N}_1=\partial {\cal S},
\ee
where $D^{\rho\mu\nu}$ is the tensorial $\delta$-function with support on ${\cal S}$ (analogous to \eqref{stringa dirac1}). We thus have
\be\label{deltac}
\tilde C_2^{\m\n}-\tilde C_1^{\m\n} =g\left(\partial^\mu \tilde D^\nu-\partial^\nu\tilde  D^\mu\right),
\ee
where $\tilde D^\mu=\frac 16\,\varepsilon^{\nu\rho\sigma\mu}D_{\nu\rho\sigma}$, i.e. the Hodge dual of the three-tensor $D^{\rho\mu\nu}$ defined in \eqref{poinc}. Since the field strength \eqref{def of F* for monopoles} must be invariant under a change of Dirac strings, we  require the four-potential to transform as follows when passing from one choice of the string to another
\be
A^\m \rightarrow A^\m  + g \tilde D^\mu.
\ee
Therefore, a finite change of the Dirac string only affects  the minimal-interaction term of the action \eqref{action of mod modmax with monopoles} which changes as follows (setting $j^\mu_e= ej^\mu$)
\be
\Delta I=\Delta \left(\int d^4x\;j_e^{\mu}A_{\mu}\right)=eg\,\int j^\mu \tilde D_\mu d^4x.
\ee
According to the theory of ``integer forms" \cite{chern2012differentiable},  the integral on the r.h.s. is an integer $n$, counting the number of intersections of the particle worldline with the three-volume ${\cal S}$.\footnote{In this derivation we put aside the issue of the {\it Dirac veto} discussed in detail in \cite{Lechner:1999ga}.} We thus have
\be\label{dI}
 \Delta I= eg\, n.
\ee
In the quantum theory the functional integrals always contain a factor $e^{\ii I}$. From \eqref{dI} we see that  $e^{\ii I}$ is Dirac string independent iff
\be\label{Diracq}
eg=2\pi n\, \qquad n=0,\pm 1, \pm 2,\ldots
\ee
which is the famous Dirac quantization condition.

Let us now derive the equations of motion of $A_\mu(x)$, $z^\mu(\tau)$ and $y^\mu(\tau)$  from the variation of \eqref{action of mod modmax with monopoles}. Varying the action with respect to $A^{\mu}$ we get the electromagnetic field equations
\begin{equation}\label{Gj}
\partial_{\mu}G^{\mu\nu}=j_e^{\nu}
\end{equation}
accompanied by the ``Bianchi's identities" sourced by the magnetic current (\ref{identita bianchi con carica senza riscaling}).

The monopole equation of motion is obtained by varying the action with respect to $z^{\mu}(\tau)$, taking into account the dependence of $w^\mu$ on $z^\mu$ in \eqref{stringa dirac}
\begin{equation}
    \dfrac{dp_g^{\mu}}{d\tau}=g\,\tilde G^{\mu\nu}\dfrac{dz_{\nu}}{d\tau}\,.
    \label{lorenz magnets}
\end{equation}

Finally, let us show that the equation of motion of the electrically charged particle derived from the action \eqref{action of mod modmax with monopoles} describes the Lorentz force
\eqref{lorentz force charges in monopoles} acting on the particle with $F^{\mu\nu}$ given in \eqref{def of F* for monopoles}. To this end we perform the variation of the relevant terms in \eqref{action of mod modmax with monopoles} with respect to $y_\mu(\tau)$ in the following way
\bea\label{onehalf1pre}
\delta_y\left( \int  j_e^\nu A_\nu \,d^4x - m_e\int d\tau\right)
=\int\left(\frac{dp_e^\mu}{d\tau}- e
\left(\partial^\mu A^{\nu} -\partial^\nu A^{\mu}\right)
v_{\nu}\right)\delta y_{\mu}d\tau\nonumber\\
= \int\left(\frac{dp_e^\mu}{d\tau}- e
\left(\partial^\mu A^{\nu} -\partial^\nu A^{\mu}-\tilde C^{\mu\nu}\right)
v_{\nu}\right)\delta y_{\mu}d\tau-e\int \tilde C^{\mu\nu}v_\nu\delta y_{\mu}d\tau\nonumber\\
= \int\left(\frac{dp_e^\mu}{d\tau}- e\,F^{\mu\nu}
v_{\nu}\right)\delta y_{\mu}d\tau-\delta_y \left(\frac{1}{2} \int d^4x\,  \tilde C_{\mu\nu}\, C^{\mu\nu}_{e}\right)\,
\eea
where $C_e^{\mu\nu}$ is associated with a formal Dirac string worldsheet attached to the electric current, namely
\be\label{Ce}
C^{\mu\nu}_{e}(x)=-e\iint d\tau d\sigma \left(\dfrac{\partial w_e^{\mu}}{\partial \tau}\dfrac{\partial w_e^{\nu}}{\partial \sigma}-\dfrac{\partial w_e^{\nu}}{\partial \tau}\dfrac{\partial w_e^{\mu}}{\partial \sigma}\right)\delta^{(4)}\left(x-w_e(\tau,\sigma)\right)
\ee
and
\begin{equation}
    w^{\mu}(\tau,\sigma)=y^{\mu}(\tau)+u_e^{\mu}(\tau,\sigma),\qquad u_e^{\mu}(\tau,0)=0\,.
    \label{we}\,
\end{equation}
so that
\be\label{dCe}
\partial_\mu C^{\mu\nu}=j_e^\mu\,.
\ee
Note that the last term in \eqref{onehalf1pre}, namely
$$
\frac{1}{2}\,\int d^4x\,  \tilde C_{\mu\nu}\,C^{\mu\nu}_{e}
$$
is $eg$ times the number of intersections between the Dirac string worldsheets of the electric particle and the monopole. Under infinitesimal variations of the electric particle worldline it is zero.\footnote{In the derivation described in \eqref{onehalf1pre} we have added and subtracted the term with $\tilde C^{\mu\nu}$ and argued that one of the two terms drops out from the final result. For the explanation of this peculiarity see the more rigorous treatment of \cite{Lechner:1999ga}, tackling also the issue of the Dirac veto.}

The equation of motion of the electrically charged particle is thus
\begin{equation}
\frac{dp^\mu}{d\tau}=
\, e\, F^{\mu}{}^\nu(y)\, v_\nu,
\label{lorentz force charges in monopoles}
\end{equation}
which is the standard Lorentz force \eqref{lorentz force charge} induced by the field strength (\ref{def of F* for monopoles}).

\subsection{ModMax coupled to dyons}

In this Section we consider an action that describes the ModMax theory coupled to an arbitrary system of dyons with masses $m_r$, world lines $y_r$ and (electric and magnetic) charges $(e_r,g_r)$, $r=1,2...,N$. This action is
\be\label{itot}
I[A,y]=\int\left(-\frac{1}{4}\cosh \gamma F^{\m\n} F_{\m\n}+\frac{1}{4}  \sinh \gamma \sqrt{(F^{\m\n} F_{\m\n})^2+ (F^{\m\n}\widetilde  F_{\m\n})^2} -A_\mu J^\mu_e  \right)d^4x
-\sum_r m_r\int ds_r.
\ee
In this case the field strength $F^{\mu\nu}$ is still given by \eqref{def of F* for monopoles}, but the $\delta$-functions associatied with the $r$ Dirac strings are now given by
\be
C^{\m\n}=\sum_r  g_r\, C_{r}^{\mu\nu}\,, \qquad C_{r}^{\mu\nu}=-\iint d\tau d\sigma \left(\dfrac{\partial w_r^{\mu}}{\partial \tau}\dfrac{\partial w_r^{\nu}}{\partial \sigma}-\dfrac{\partial w_r^{\nu}}{\partial \tau}\dfrac{\partial w_r^{\mu}}{\partial \sigma}\right)\delta^{(4)}\left(x-w_r(\tau,\sigma)\right)\,.
\ee
The total electric current is
\be
J^\mu_e=\sum_r e_r j_r^\mu, \qquad j_r^\mu=\int d\tau\; \delta^{(4)}\left(x-z_r(\tau)\right)\dfrac{dz_r^{\mu}}{d\tau}\,.
\ee
The electromagnetic field equations become
\begin{equation}\label{MMdyon}
  \partial_{\mu}G^{\mu\nu}=J_e^{\nu}\,\qquad \partial_{\mu}\tilde F^{\mu\nu}=J_g^{\nu}
\end{equation}
instead of \eqref{identita bianchi con carica senza riscaling} and \eqref{Gj}, where $J^\mu_g=\sum_r g_r\, j_r^\mu$ and $G^{\mu\nu}$ was defined in \eqref{G}. And the equations of motion of the dyons are
\begin{equation}
    \dfrac{dp_r^{\mu}}{d\tau_r}=\left(e_r\, F^{\mu\nu}+g_r\,\tilde G^{\mu\nu}\right)\dfrac{dy_{r\nu}}{d\tau_r}\,.
    \label{lorentz force for dioni}
\end{equation}
Note that the equations of motion \eqref{MMdyon} and \eqref{lorentz force for dioni} are formally invariant under the duality rotations \eqref{Duality} if we assume that the electric and magnetic charges of the each dyon get transformed accordingly
\be\label{egDuality}
 \begin{pmatrix}e'\\
 g'\end{pmatrix}
=
\begin{pmatrix} \cos\alpha & \sin\alpha\\
 -\sin\alpha & \cos\alpha \end{pmatrix}\,
\begin{pmatrix}e\\
 g\end{pmatrix}\,.
\ee
 Under a different choice of the Dirac strings (see \eqref{deltac}) accompanied by the shift of the four-vector potential
\be\label{dirac1}
C^{\mu\nu}_1\,\rightarrow \,C^{\mu\nu}_2=C_1^{\mu\nu}+ \sum_r g_r\left(\partial^\mu \tilde D^\nu_r-\partial^\nu\tilde  D^\mu_r\right),\qquad
A^\m \rightarrow A^\m  + \sum_r g _r\tilde D^\mu_r,
\ee
the action changes as follows
\be\label{vari}
\Delta I=\Delta \left(\int d^4x\;J_e^{\mu}A_{\mu}\right)=\sum_{r,s} e_rg_s \int j_{r\mu}\,\tilde D^{\mu}_sd^4x=\sum_{r,s} e_rg_s N_{rs},\quad N_{rs}\in\mathbb{N},
\ee
which leads to the Dirac-quantization conditions
\be\label{Diracq1}
e_rg_s=2\pi n_{rs}, \quad n_{rs}=0,\pm 1, \pm 2,\ldots,
\ee
for each pair $(r,s)$, that generalize the condition \eqref{Diracq}.

Notice that the conditions \eqref{Diracq1} violate the $SO(2)$ duality invariance \eqref{egDuality}. They are only invariant under the discrete $Z_4$ transformations generated by $e_r \to g_r$, $g_r\to - e_r$ (and the same for the $s$-dyon) and $n_{rs} \to - n_{sr}$. To get an $SO(2)$-invariant charge quantization condition, following \cite{Lechner:1999ga} we add to the action \eqref{itot} a term involving only the Dirac strings, which does not change the equations of motion of the electromagnetic field, nor those of the dyons. The modified action takes the form
\be\label{iprimo}
I'=I+\frac{1}{4}\int\sum_{r,s} e_r\,\tilde C_{r}^{\mu\nu}\,g_s\,C_{s\mu\nu}.
\ee
The dependence of the second term in \eqref{iprimo} on the particle trajectories is proportional to an integer multiple of $n\pi$, and hence does not contribute to infinitesimal variations of the classical action with respect to $z_r^\mu(\tau)$.

Under the change of the choice of the Dirac strings  \eqref{dirac1}, with the use of \eqref{dCj} and \eqref{vari}, one finds that the variation of the modified action is
\be
\Delta I'=\Delta I-\frac{1}{2}\sum_{r,s}\left(e_rg_s+e_sg_r\right)\int j_\mu^r \tilde D^{\mu}_s\, d^4x=\frac{1}{2}\sum_{r,s}\left(e_rg_s- e_sg_r\right)\int j_\mu^r \tilde D^{\mu}_s\,d^4x=\frac{1}{2}\sum_{r,s}\left(e_rg_s- e_sg_r\right) N_{rs},
\ee
Requiring that this variation does not change $e^{iI}$ in the functional integrals we get the Schwinger quantization condition
\be\label{schwinger}
\frac{1}{2}\left(e_rg_s-e_sg_r\right)=2\pi  n_{rs}, \qquad n_{rs}=0,\pm 1, \pm 2,\ldots.
\ee
which is invariant under the $SO(2)$ rotation \eqref{egDuality} of the charges.

But there is even more, as was shown in \cite{Lechner:1999ga} for Maxwell's theory, by adding the same second term as in \eqref{iprimo} the resulting effective action for dyons obtained by performing the functional integral over the gauge field $A^\mu$ is invariant under $SO(2)$ duality. We assume that also the ModMax theory coupled to dyons and based on the modified action \eqref{iprimo}, will eventually be invariant under $SO(2)$ duality.

\subsection{The Lienard-Wiechert field of a single dyon}
Let us now consider the coupling to ModMax of a single dyon carrying the charges $(e,g)$, i.e. we take $r=1$ in \eqref{itot}-\eqref{lorentz force for dioni}.
The electromagnetic field equations have the same form as in \eqref{MMdyon}.
We will now show that for the single dyon the equations \eqref{MMdyon} are satisfied by the Lienard-Wiechert fields induced by the moving dyon.  As the ansatz we take the following generalization of the Lienard-Wiechert field strength \eqref{LWe}
\begin{equation}
    F_{dLW}^{\mu\nu}=e^{-\gamma}\,e\;f^{\mu\nu}-g\;\tilde f^{\mu\nu}\,,
    \label{dion field sol}
\end{equation}
where $f^{\mu\nu}$ has been defined in \eqref{LW fields} and is such that
\be\label{ftf1}
\partial_\mu f^{\mu\nu}=\int d\tau\; \delta^{(4)}\left(x-y(\tau)\right)\dfrac{dy^{\nu}}{d\tau}\,,\qquad \partial_\mu \tilde f^{\mu\nu}=0\,, \qquad f_{\mu\nu}\,\tilde f^{\mu\nu}=0.
\ee
So the field strength \eqref{dion field sol} is such that
\be\label{dF=j}
\partial_\mu F_{dLW}^{\mu\nu}=e^{-\gamma}\,j_e\,, \qquad \partial_\mu \tilde F_{dLW}^{\mu\nu}=j_g.
\ee
So the second equation in \eqref{MMdyon} is also satisfied and it remains to show that the above relations imply that the first equation is satisfied as well. To this end we should compute the form of $G^{\mu\nu}$ in \eqref{G} for the ansatz \eqref{dion field sol}. Using the last equation in \eqref{ftf1} we find that \footnote{A recent result obtained in ModMax for a charged black hole in a Rindler frame \cite{Barrientos:2022bzm} is a particular case of the  general solution considered here.}
\be\label{SPLW}
 S=-\frac 14(e^{-2\gamma}\,e^2-g^2)\;f_{\mu\nu}f^{\mu\nu}\,,\quad P=-\frac{e^{-\gamma}}2\,eg\;f_{\mu\nu}f^{\mu\nu}\quad\implies\quad P=\dfrac{2e^{-\gamma}\,eg}{e^{-2\gamma}\,e^2-g^2}S\,
\ee
and hence
\be\label{GLW}
G^{\mu\nu}=e\;f^{\mu\nu}-e^{-\gamma}g\;\tilde f^{\mu\nu}\,,
\ee
which satisfies the first equation in \eqref{MMdyon}. Thus we have proved that the  Lienard-Wiechert fields created by the dyon are indeed exact solutions of the ModMax equations of motion. However, in general,  linear combinations of Lienard-Wiechert fields created by several dyons will not be solutions of the ModMax non-linear field equations.

    The Lorentz force of the Lienard-Wiechert fields acting on a test dyon of an electric charge $q$ and a magnetic charge $r$ has the form (see \eqref{lorentz force for dioni} with $r=1$,  and use \eqref{GLW} and \eqref{dion field sol})
\be\label{LFLWeg}
\dfrac{dp^{\mu}}{d\tau}=\left[e^{-\gamma}(qe+rg)\;f^{\mu\nu}+(re-qg)\;\tilde f^{\mu\nu}\right]{v}_{\nu}\,.
\ee
As has already been observed for the Coulomb force \eqref{Coulomb},  we see that in the ``electric" part of the ModMax theory (first term on the l.h.s. of \eqref{LFLWeg}), the Lorentz force differs from that in Maxwell's theory by a factor of $e^{-\gamma}$, while the ``magnetic" parts are the same (the second term in \eqref{LFLWeg}). We will now show that the electric parts of the Lorentz forces in the two theories can be made equal by a suitable rescaling of the ModMax Lagrangian. But this will lead to a difference in the magnetic part of the Lorentz forces.

\section{Different scalings of the ModMax Lagrangian}\label{difs}
 Let us consider the ModMax Lagrangian rescaled with the $e^{-\gamma}$ factor \footnote{As was noticed in the Note Added of \cite{Bandos:2020hgy}, a different form of this Lagrangian was somewhat implicitly present in \cite{Denisova:2019lgr} among a more general class of conformally invariant models of non-linear electrodynamics.}
\begin{equation}
    \mathcal{L}= e^{-\gamma}(\cosh{\gamma}\;S+\sinh{\gamma}\sqrt{S^2+P^2})
    \label{egammaL}
\end{equation}
Then
\begin{equation}
    G^{\mu\nu}=-\dfrac{2\,\partial \mathcal{L}}{\partial F_{\mu\nu}}= e^{-\gamma}\left[\left(\cosh{\gamma}+\sinh{\gamma}\dfrac{S}{\sqrt{P^2+S^2}}\right)F^{\mu\nu}+\left(\sinh{\gamma}\dfrac{P}{\sqrt{P^2+S^2}}\right)\text{ }\tilde F^{\mu\nu}\right]
    \label{def Gmunu nuova}
\end{equation}
and now for $P=0$ and $S>0$ we have $G^{\mu\nu}=F^{\mu\nu}$ and the field equations reduce to the linear Maxwell equations. These have the standard Lienard-Wiechert fields \eqref{LW fields} as solutions, while the Lorentz force remains the same as in (\ref{lorentz force charge}). Therefore now the Coulomb law takes exactly the same form as in Maxwell's theory.

Let us note that the rescaling of the ModMax Lagrangian results in the following modification of the duality-invariance condition (\ref{Dualitycond})
\begin{equation}
G_{\mu\nu} \tilde G^{\mu\nu}=e^{-2\gamma}F_{\mu\nu}
\tilde F^{\mu\nu}
\label{nuova condizione dualita}
\end{equation}
and so the duality transformation (\ref{Duality}) now involves the rotation of $(G^{\mu\nu},e^{-\gamma}\,\tilde F^{\mu\nu})$  instead of $(G^{\mu\nu}, \tilde F^{\mu\nu})$.
Hence, if we want to couple the Lagrangian density \eqref{egammaL} to a dyon in such a way that the formal $SO(2)$ duality invariance of the equations of motion holds for the electric and magnetic charges transformed as in \eqref{egDuality}, the field equations \eqref{MMdyon} should be modified as follows
\be\label{MMegm}
\partial_{\mu}G^{\mu\nu}=j_e^{\nu}\,,\qquad    e^{-\gamma}\partial_{\mu}\tilde F^{\mu\nu}=j_g^{\nu}\,.
\ee
With this choice of the form of the second equation above, the tensor distribution $C^{\mu\nu}$ in \eqref{def of F* for monopoles} acquires the factor $e^{\gamma}$ in comparison with \eqref{stringa dirac}, namely
\begin{equation}
    C^{\mu\nu}(x)=-e^{\gamma}g\iint d\tau d\sigma \left(\dfrac{\partial w^{\mu}}{\partial \tau}\dfrac{\partial w^{\nu}}{\partial \sigma}-\dfrac{\partial w^{\nu}}{\partial \tau}\dfrac{\partial w^{\mu}}{\partial \sigma}\right)\delta^{(4)}\left(x-w(\tau,\sigma)\right)\,.
    \label{stringa dirac}
\end{equation}
The above rescalings then imply that the Lienard-Wiechert fields of the dyon that satisfy \eqref{MMegm} have the following form
\begin{equation}\label{LWdm}
    F^{\mu\nu}=e\;f^{\mu\nu}-e^{\gamma}\,g\;\tilde f^{\mu\nu}\,,\qquad  G^{\mu\nu}=e\;f^{\mu\nu}-e^{-\gamma}\,g\;\tilde f^{\mu\nu}\,.
\end{equation}
These fields produce the following Lorentz force on a test dyon carrying the charges  $(q,r)$
\be\label{LFLWresc}
\dfrac{dp^{\mu}}{d\tau}=\left[(qe+rg)\;f^{\mu\nu}+e^{\gamma}(re-qg)\;\tilde f^{\mu\nu}\right]{v}_{\nu}\,.
\ee

The ``electric" part of the force is the same as in Maxwell's theory, while the ``magnetic" part differs by the factor of $e^{\gamma}$. Note also that the Dirac quantization condition \eqref{Diracq} and the Schwinger quantization condition \eqref{schwinger}  acquire now the factor of $e^{-\gamma}$
\be\label{ds}
qg=2\pi\,n\,e^{-\gamma}\,, \qquad \frac 12(re-qg)= 2\pi\,n\,e^{-\gamma}\,.
\ee
One can remove the factors of $e^{-\gamma}$ from the equations \eqref{MMegm}, the ``magnetic" part  of \eqref{LFLWresc} and the quantization conditions \eqref{ds} by redefining the physical values of the magnetic charges $g\to e^{-\gamma}\,g$ and $r\to e^{-\gamma}\,r$. Then the Lorentz force takes the form
\be\label{LFLWresc1}
\dfrac{dp^{\mu}}{d\tau}=\left[(qe+e^{-2\gamma}rg)\;f^{\mu\nu}+(re-qg)\;\tilde f^{\mu\nu}\right]{v}_{\nu}\,,
\ee
which preserves Schwinger's quantization condition \eqref{schwinger}.

In any case, from \eqref{LFLWresc} and \eqref{LFLWresc1} it follows that via the Lorentz force one can distinguish ModMax described by the Lagrangian \eqref{egammaL} from Maxwell's theory only in the presence of magnetic charges.\footnote{Note that the rescaling of the ModMax Lagrangian and magnetic charges considered in this Section also affects the contribution of charges to the black hole solutions in General Relativity coupled to ModMax considered in \cite{Flores-Alfonso:2020euz,Bordo:2020aoz,Flores-Alfonso:2020nnd,Amirabi:2020mzv,Bokulic:2021dtz,Zhang:2021qga,Bokulic:2021xom,Ali:2022yys,Kruglov:2022qag,Ortaggio:2022ixh,Barrientos:2022bzm}.}
This suggests to look for other physical effects in which ModMax non-linearity is manifested independently of the rescaling of its Lagrangian and charges. One of these effects is the birefringence of light propagating in a uniform strong electromagnetic background computed for ModMax in \cite{Bandos:2020jsw,Flores-Alfonso:2020euz}. Another one is the Compton effect which we will analyze in the next Section.

\section{Compton effect in ModMax}\label{Compton}
The Compton effect is a well known photon-electron scattering process. It consists in the change in the change of the  wavelength of the photon when scattered by a free electron and manifests the corpuscular nature of light.
In \cite{Neves:2021tbt} properties of Compton scattering were derived in a generic theory of non-linear electrodynamics (analytic in $S$ and $P$) in a uniform magnetic background and used to study features of this effect in some concrete models, such as the Heisenberg-Euler, Born-Infeld and a logarithmic electrodynamics. Like in the case of birefringence, switching on a strong uniform electromagnetic background allows to distinguish the Compton effect in a given non-linear electrodynamics model from that in Maxwell's theory. In this Section we will compute the Compton effect in ModMax in the presence of a magnetic background.

 To this end we take the dispersion relations for the 4d wave vector $ k^\mu=(\omega,\mathbf k)$ of an electromagnetic wave propagating in a uniform magnetic background $\mathbf B$ computed for ModMax in \cite{Bandos:2020jsw}.  The wave  undergoes birefringence and splits in two rays. One of them propagates along the standard relativistic light cone
\be\label{rela}
 \omega^2=|\mathbf k|^2,
\ee
 while the other one propagates along a path characterized by the following dispersion relation
\begin{equation}
    \omega^2=|\mathbf{k}|^2\left(\cos^2{\beta}+e^{-2\gamma}\sin^2{\beta}\right),
 \label{omega k wave B}
\end{equation}
where
\be\label{kBangle}
    \cos{\beta}=\dfrac{\mathbf{k}\cdot\mathbf{B}}{|\mathbf{k}||\mathbf{B}|}
\ee
and $\beta$ is the angle between the direction of the light ray and the background magnetic field. In this case (for $\beta\not = 0$ and $\gamma>0$) the velocity of propagation $v=\sqrt{\left(\cos^2{\beta}+e^{-2\gamma}\sin^2{\beta}\right)}$ is less than the speed of light in the vacuum ($c=1$).

We are interested in the Compton effect of the photon obeying eq. \eqref{omega k wave B}, because the Compton scattering of the relativistic photon \eqref{rela} will be the same as in Maxwell's theory. Since we are using natural units in which $c=1$ and $\hbar=1$, the dispersion relation \eqref{omega k wave B} is straightforwardly promoted to the condition on the four-momentum $p^\mu=(E,\mathbf p)$ of the photon
\begin{equation}
    E^2=|\mathbf{p}|^2\left(\cos^2{\beta}+e^{-2\gamma}\sin^2{\beta}\right)\qquad\text{with}\qquad \cos{\beta}=\dfrac{\mathbf{p}\cdot\mathbf{B}}{|\mathbf{p}||\mathbf{B}|}\,.
    \label{EP rel Compton}
\end{equation}

\subsection{Compton effect in the magnetic field orthogonal to the incoming photon momentum}\label{perp}
As in \cite{Neves:2021tbt} we first consider a magnetic field orthogonal to the direction of the incoming photon. We assume that  an experimental set-up is such that the electron can be considered at rest before scattering. Hence, the initial four-momentum of the electron is $p^\mu_e=(m_e,0)$, where $m_e=\lambda_e^{-1}$  are the electron mass and the Compton wavelength, respectively. Upon scattering with the photon the electron acquires the four momentum $p_e^{\mu}=(E_e,\mathbf p_e)$.  The four-momentum of the incoming photon is $p^\mu=(E,\mathbf{p})$ and that of the outgoing one is $p'{}^\mu=(E',\mathbf{p}')$.
Because of our choice of the orientation of $\mathbf{B}$  ($\mathbf{p}\cdot\mathbf{B}=0$), from \eqref{EP rel Compton} we have
\begin{equation}
    E^2=e^{-2\gamma}\,|\mathbf{p}|^2\,, \qquad  E'^2=|\mathbf{p'}|^2\left(\cos^2{\beta'}+e^{-2\gamma}\sin^2{\beta'}\right)\equiv |\mathbf{p'}|^2\,f^2(\beta',\gamma)
    \label{epe'p'},
\end{equation}
where $\beta'$ is the angle between $\mathbf B$ and the momentum $\mathbf p'$ of the outgoing photon.

Conservation of the $4$-momentum implies that
\begin{equation}
    E_e^2=(E+m-E')^2\,,\qquad \mathbf{p}_e^2=(\mathbf{p}-\mathbf{p}')^2=|\mathbf{p}|^2+|\mathbf{p}'|^2-2\mathbf{p}\cdot\mathbf{p}'\,.
\end{equation}
Remembering that $E_e^2-\mathbf p_e^2=m_e^2$ and using (\ref{epe'p'}) we get the following relation
\begin{equation}
    p^2(e^{-2\gamma}-1)+p'^2(f^2-1)-2 p p'(e^{-\gamma}\,f-\cos{\theta})+2m\,(e^{-\gamma}p-p'\,f)=0\,,
    \label{}
\end{equation}
where $p=|\mathbf{p}|$, $p'=|\mathbf{p}'|$ and $\theta$ is the angle between $\mathbf{p}$ and $\mathbf{p}'$.

For the wavelength $\lambda=1/p$ and $\lambda'=1/p'$ of the incoming and outgoing photon the above equation takes the following form
\begin{equation}
    \lambda'^2(e^{-2\gamma}-1)+\lambda^2(f^2-1)-2 \lambda\lambda'(e^{-\gamma}\,f-\cos{\theta})+2\dfrac{\lambda\lambda'}{\lambda_e}\,(e^{-\gamma}\lambda'-\lambda\,f)=0\,.
\end{equation}
Solving this equation for positive $\lambda'$ we find
\begin{equation}
    \lambda'=\dfrac{1}{2-2\sinh{\gamma}\,(\lambda_e/\lambda)}\bigg\{f\lambda+\lambda_e(f-\cos{\theta})+
    \label{lambda1 B ort2}
\end{equation}
\[
+\sqrt{[e^{\gamma}f\lambda+\lambda_e(f-e^{\gamma}\cos{\theta})]^2+2e^{\gamma}\lambda\lambda_e(1-f^2) (1-\sinh{\gamma}\,\lambda_e/\lambda)}\,\bigg\}\,.
\]
One can see that for $\gamma=0$  we have $f=1$ and \eqref{lambda1 B ort2} reduces to the usual Compton scattering formula, as expected.

The above expression is rather complicated, but assuming that $\gamma$ is small one can perform the Taylor expansion of (\ref{lambda1 B ort2}) up to the first order in $\gamma$ which results in the following simpler expression
\be\label{delta lambda B ort}
    \lambda' - \lambda = \lambda_e(1-\cos{\theta})+\gamma\lambda\left[1-\dfrac{\sin^2{\beta'}}{1+\lambda_e/\lambda\,(1-\cos{\theta})}\right]\left[1+\left(\dfrac{\lambda_e}{\lambda}+\dfrac{\lambda_e^2}{\lambda^2}\right)(1-\cos{\theta})\right]+ O(\gamma^2)\,.
\ee
In ModMax the effect differs from that in Maxwell's theory by
\bea
    \Delta\lambda &\equiv& \lambda' - \lambda-\lambda_e(1-\cos{\theta})=
    \gamma\lambda\left[1-\dfrac{\sin^2{\beta'}}{1+\lambda_e/\lambda\,(1-\cos{\theta})}\right]\left[1+\left(\dfrac{\lambda_e}{\lambda}+\dfrac{\lambda_e^2}{\lambda^2}\right)(1-\cos{\theta})\right]+ O(\gamma^2)\,\nonumber\\
    &=&\gamma\left[\dfrac{\lambda\,\cos^2{\beta'}}{1+\lambda_e/\lambda\,(1-\cos{\theta})}+\dfrac{\lambda_e(1-\cos{\theta})}{1+\lambda_e/\lambda\,(1-\cos{\theta})}\right]\left[1+\left(\dfrac{\lambda_e}{\lambda}+\dfrac{\lambda_e^2}{\lambda^2}\right)(1-\cos{\theta})\right]+O(\gamma^2)\,.
    \label{lambda' Bort new}
\eea
We see that, with the chosen direction of the background magnetic field, in ModMax the difference of the wavelengths of the incoming and outgoing photons is a bit greater than that in Maxwell's electrodynamics. This can be attributed to two contributions. The first one is proportional to $\cos^2\beta'$ in \eqref{lambda' Bort new} and is due to the refraction in the chosen magnetic background which is such that the effective velocity  of the outgoing photon $v=\sqrt{\cos^2{\beta'}+e^{-2\gamma}\sin^2{\beta'}}\geq e^{-\gamma}$  (read from \eqref{epe'p'}) is in general greater than that of the incoming one. This contribution increases for larger wavelengths of the incoming photon.  And the second contribution is non-zero even in the case in which $\beta' = \pi/2$ and the velocity does not change but is still less than the vacuum speed of light ($v=e^{-\gamma}$)
\be
    \Delta\lambda = \gamma\,\dfrac{\lambda_e\,(1-\cos{\theta})}{1+\lambda_e/\lambda\,(1-\cos{\theta})}\left[1+\left(\dfrac{\lambda_e}{\lambda}+\dfrac{\lambda_e^2}{\lambda^2}\right)(1-\cos{\theta})\right]+ O(\gamma^2)\nonumber\\
\ee
When the trajectory of the scattered photon is along the magnetic field $(\beta'=0)$ the relation \eqref{delta lambda B ort} simplifies to
\begin{equation}
\Delta\lambda = {\gamma\lambda}\left(1+\left(\dfrac{\lambda_e}{\lambda}+\dfrac{\lambda_e^2}{\lambda^2}\right)(1-\cos{\theta})\right)+ O(\gamma^2)\,.
\end{equation}

\subsection{Compton effect in the magnetic field parallel to the incoming photon momentum}\label{B //}
One can consider a different setup for studying the Compton scattering by orienting the uniform magnetic field $\mathbf{B}$ along the momentum $\mathbf{p}$ of the incoming photon. With this configuration, indicating as before with $\theta$ the angle between $\mathbf{p}$ and $\mathbf{p}'$ (which is now the same as the angle between $\mathbf{B}$ and $\mathbf{p}'$ ($\theta=\beta'$)), we see that the energy-momentum relations (\ref{EP rel Compton}) take the following form
\begin{equation}
    E^2=\mathbf{p}^2\,,\qquad E'^2=\mathbf{p}'^2 f^2(\gamma,\theta)\quad \text{where}\quad f^2(\gamma,\theta)=\cos^2{\theta}+e^{-2\gamma}\sin^2{\theta}\,.
    \label{e-p eq B //}
\end{equation}
Imposing the conservation of the four-momentum and proceeding as above we find the relation
\begin{equation}
    \lambda'-f\lambda-\dfrac{\lambda\lambda_e}{2\lambda'}(1-f^2)-\lambda_e f+\lambda_e\cos{\theta}=0\,.
\end{equation}
Solving this equation for positive $\lambda'$ we get
\begin{equation}
    \lambda' = \dfrac{1}{2}f\lambda + \dfrac{1}{2}\lambda_e(f-\cos{\theta})+\dfrac{1}{2}\sqrt{\lambda_e^2(f-\cos{\theta})^2+\lambda^2f^2+2\lambda\lambda_e(1-f\cos{\theta})}\,.
\end{equation}
The Taylor expansion of this expression up-to the first order in $\gamma$  gives
 \begin{equation}
 \Delta\lambda=  \lambda'- \lambda - \lambda_e(1-\cos{\theta})=-\gamma\,\dfrac{\lambda\,\sin^2{\theta}}{1+\lambda_e/\lambda(1-\cos{\theta})}\left[1+\left(\dfrac{\lambda_e}{\lambda}+\dfrac{\lambda_e^2}{\lambda^2}\right)(1-\cos{\theta})\right]+O(\gamma^2)
    \label{delta lambda B para}
\end{equation}
In contrast to the case discussed in Section \ref{perp} we see that the difference of the wavelengths of outgoing and incoming photons is smaller than in Maxwell's electrodynamics. This can be attributed to the fact that, as can be deduced from \eqref{e-p eq B //}, the effective velocity of the outgoing photon is less than that of the incoming one due to the refraction in the magnetic background.  In general, the difference between the effects in the two theories increases for larger wavelengths, but when the photon is recoiled, i.e. $\theta=\pi$, there is no difference with Maxwell's theory, because in this case the incoming and the recoiled photons travel with the speed of light in the vacuum.

\section{Conclusion}
In this paper we have studied the interaction of electrically and magnetically charged particles in the ModMax theory. We have found that the Lienard-Wiechert fields induced by a moving electric or magnetic particle, or a dyon are exact solutions of the ModMax equations of motion (with appropriately rescaled charges). The Lorentz force, and in particular the Coulomb law between two electric particles, can be made the same as in Maxwell's theory by choosing a suitable definition of the physical electric charges (which corresponds to a certain rescaling of the ModMax Lagrangian). However, in the presence of magnetic charges  the Lorentz force in ModMax always differs from that in Maxwell's theory independently of the rescaling of the charges. Different rescalings with functions of the ModMax parameter $\gamma$ result in different values of the permittivity and the permeability in ModMax in comparison with their vacuum values in Maxwell's theory. This can presumably be used for mimicking properties of some materials. We have then considered physical effects, namely birefringence and the Compton scattering, which manifest the discrepancy of ModMax from Maxwell's theory independently of the scaling of the ModMax Lagrangian and the definition of the physical charges. It will be of interest to study other physical effects, e.g. light-by-light scattering and the Hall effect as a measure of the fine structure constant in this theory. We have also presented alternative forms of the ModMax Lagrangian constructed with the use of auxiliary scalar fields. These forms may be useful for understanding the origin of ModMax as an effective field theory and for its quantization.
\\

\noindent
{\Large \bf {Acknowledgements}} 

\medskip
\noindent
The authors are grateful to Igor Bandos, Gregory Korchemsky and Paul Townsend for interest to this work, useful discussions and remarks.
Work of K.L, P.M. and D.S. has been partially supported by the INFN Research Project ``String Theory and Fundamental Interactions''
(STEFI). D.S. was also partially supported by the Spanish MICINN/FEDER (ERDF EU) grant PGC2018-095205-B-I00 and by the Basque Government Grant IT-979-16.

%\newpage

\if{}
\bibliographystyle{abe}
\bibliography{references}{}
\fi

\providecommand{\href}[2]{#2}\begingroup\raggedright\endgroup

\end{document}